\magnification\magstep1
\font\BBig=cmr10 scaled\magstep2
\font\small=cmr7


\def\title{
{\bf\BBig
\centerline{
Non-relativistic Maxwell-Chern-Simons Vortices}
\bigskip
}
} 


\def\authors{
\centerline{
M. HASSA\"INE\foot{e-mail: hassaine@univ-tours.fr},
P.~A.~HORV\'ATHY\foot{e-mail: horvathy@univ-tours.fr}
 and
J.-C.~YERA\foot{e-mail: yera@univ-tours.fr}}
\bigskip
\centerline{
D\'epartement de Math\'ematiques}
\medskip
\centerline{Universit\'e de Tours}
\medskip
\centerline{Parc de Grandmont,
F--37200 TOURS (France)
}
}

\def\runningauthors{
Hassa\"\i ne, Horv\'athy \&
Yera
}

\def\runningtitle{
Non-relativistic Maxwell-Chern-Simons
}


\voffset = 1cm 
\baselineskip = 13pt 

\headline ={
\ifnum\pageno=1\hfill
\else\ifodd\pageno\hfil\tenit\runningtitle\hfil\tenrm\folio
\else\tenrm\folio\hfil\tenit\runningauthors\hfil
\fi
\fi}

\nopagenumbers
\footline={\hfil} 


\def\and{\qquad\hbox{and}\qquad}
\def\where{\qquad\hbox{where}\qquad}

\def\kikezd{\parag\underbar} 

\def\IR{{\bf R}}
\def\smallover#1/#2{\hbox{$\textstyle{#1\over#2}$}}
\def\2{{\smallover 1/2}}
\def\ccr{\cr\noalign{\medskip}} 
\def\parag{\hfil\break} 
\def\={\!=\!}
\def\p{\partial}

\def\L{{\cal L}}
\def\I{{\rm I}}


\newcount\ch 
\newcount\eq 
\newcount\foo 
\newcount\ref 

\def\chapter#1{
\parag\eq = 1\advance\ch by 1{\bf\the\ch.\enskip#1}
}

\def\equation{
\leqno(\the\ch.
\the\eq)\global\advance\eq by 1
}

\def\foot#1{
\footnote{($^{\the\foo}$)}{#1}\advance\foo by 1
} 

\def\reference{
\parag [\number\ref]\ \advance\ref by 1
}

\ch = 0 
\foo = 1 
\ref = 1 


\title
\vskip10mm
\authors
\vskip.20in

\parag{\bf Abstract.}
{\it The non-relativistic Maxwell-Chern-Simons model recently
introduced by Manton is shown to admit
self-dual vortex solutions with  non-zero electric field. 
The interrelated ``geometric'' and ``hidden''
symmetries are explained.
The theory is also extended to (non-relativistic) spinors.
A relativistic, self-dual model, whose non-relativistic limit is
the Manton model is also presented.
The relation to previous work  
is discussed.}

\vskip15mm
\noindent
(\the\day/\the\month/\the\year)
\bigskip
\medskip\noindent
{\sl Annals of Physics} (N. Y.). (to be published)
\vskip35mm

\vskip5mm
\vfill\eject

\chapter{Introduction}

In a recent paper [1], Manton proposed a modified version
of the Landau-Ginzburg model for describing
Type II superconductivity.
His Lagrange  density is a subtle mixture blended from
 the standard Landau-Ginzburg expression, augmented with
the Chern-Simons term:
$$
\eqalign{
{\cal L}=
&-{1\over2}B^2+\gamma{i\over2}\big(\phi^*D_t\phi-\phi(D_t\phi)^*\big)
-{1\over2}\big|\vec{D}\phi\big|^2
-{\lambda\over8}\big(1-|\phi|^2\big)^2
\cr
&+\mu\big(Ba_t+E_2a_1-E_1a_2\big)
-\gamma a_t-\vec{a}\cdot\vec{J}^{T},
\cr}
\equation
$$
where $\mu$, $\gamma>0$, $\lambda>0$ are constants,
$D_t\phi=\partial_t\phi-ia_t\phi$,
$D_i\phi=\partial_i\phi-ia_i\phi$,
$B=\partial_1a_2-\partial_2a_1$ is the magnetic field
and
$\vec{E}=\vec\nabla a_t-\partial_t\vec{a}$
is the electric field.
This Lagrangian  differs from the standard expression
in that 
(i) it is linear in $D_{t}\phi$; 
(ii) the electric term $\vec{E}^2$ is missing; 
(iii) it  includes the terms
$-\gamma a_{t}$ and
$-\vec{a}\cdot\vec{J}^{T}$, where $\vec{J}^{T}$
is the (constant) transport current. 
The properties (i) and (ii) come from the requirement of Galilean
rather than Lorentz invariance [2].
The term $-\gamma a_{t}$ results in modifying
the Gauss law (eqn. (1.4) below);
the term $-\vec{a}\cdot\vec{J}^{T}$
is then needed in order to restore the Galilean invariance.
To be so, the transport current has to transform as
 $\vec{J}^{T}\to\vec{J}^{T}+\gamma\vec{v}$
under a Galilei boost [1].

The field equations derived from (1.1) 
are 
$$
i\gamma D_t\phi=
-{1\over2}\vec{D}^2\phi
-{\lambda\over4}\big(1-|\phi|^2\big)\phi,
\equation
$$
\null\vskip-13mm
$$
\epsilon_{ij}\partial_{j}B=
J_{i}-J^{T}_{i}+2\mu\,\epsilon_{ij}\,E_j,
\equation
$$
\null\vskip-13mm
$$
2\mu B=\gamma\big(1-|\phi|^2\big),
\equation
$$
where the (super)current 
is
$
\vec{J}=({1/2i})\big(\phi^*\vec{D}\phi-\phi(\vec{D}\phi)^*\big).
$
The matter field satisfies hence a
gauged, planar non-linear Schr\"odinger equation.
The second equation is Amp\`ere's
law without the displacement current, as usual in
the ``magnetic-type'' Galilean electricity [2].
 The last equation called the Gauss law
is the (modified) ``Field-Current Identity''
of Jackiw and Pi [3]. 

Conventional Landau-Ginzburg theory admits finite-energy, static, 
 purely magnetic vortex solutions [4].
For a specific value of the coupling constant, one can find solutions
by solving instead the first-order ``Bogomolny'' equations [5],
$$\eqalign{
&(D_1+iD_2)\phi=0,
\cr
&2B=1-\vert\phi\vert^2.
\cr}
\equation
$$

\goodbreak
Now, as observed by Manton, these same solutions
yield magnetic vortices 
with $a_{t}=0$, also in his model, when
$\vec{J}^T=0$,
 $\lambda=1$ and $\mu=\gamma$.
Manton also conjectures the existence of further
solutions with a non-vanishing electric field. 

\goodbreak
In this Paper, we show that this is indeed the case~: a slight 
generalization of the self-duality equations (1.5) does indeed
provide self-dual vortices with nonzero electric field.
Being self-dual, these solutions are stable.

Next, using the equivalence of the model with
one in constant  external electric and magnetic fields,
we discuss the subtle symmetries.
In Section 4, we construct self-dual non-relativistic spinorial 
vortices along the same lines.
In Section 5 we present a self-dual, relativistic model of 
the same type, whose non-relativistic limit is the Manton model.
Finally, we compare our results to those obtained by other
people.

\goodbreak
\chapter{Self-dual vortices}

 In the frame where $\vec{J}^{T}=0$ (which 
can always be achieved by a Galilei boost),
the static Manton equations (1.2-4) read 
$$\eqalign{
&\gamma a_t\phi=
-{1\over2}\vec{D}^2\phi
-{\lambda\over4}\big(1-|\phi|^2\big)\phi,
\cr
&\vec\nabla\times B=\vec{J}+2\mu\vec\nabla\times a_t,
\cr
&2\mu B=\gamma\big(1-|\phi|^2\big).
\cr}
\equation
$$

Let us try  to solve these by the first-order Ansatz
$$\eqalign{
&(D_{1}\pm iD_2)\phi=0,
\cr
&2\mu B=\gamma\big(1-|\phi|^2\big).
\cr}
\equation
$$
From the first of these relations we infer that
$
\vec{D}^2=\mp i\big[D_1, D_2\big]=\mp B
$
and
$\vec{J}=\mp\2\vec\nabla\times\varrho,
$
where $\varrho=|\phi|^2$.
 Inserting into the non-linear Schr\"odinger equation
we find that it is identically satisfied when 
$
a_t=(\pm{1/4\mu}-{\lambda/4\gamma})(1-\varrho).
$
Then from Amp\`ere's law we get that $\lambda$ has to be
$$
\lambda=\pm2{\gamma\over\mu}-{\gamma^2\over\mu^2}.
\equation
$$
The scalar potential is thus
$$
a_t=
\smallover1/{4\mu}\big(\mp1+\smallover{\gamma}/{\mu}\big)\,
\big(1-\varrho\big).
\equation
$$

Then the vector potential is expressed using the ``self-dual'' 
(SD) Ansatz (2.2)
as
$$
\vec{a}=\pm\2\vec\nabla\times\log\varrho+\vec\nabla\omega,
\equation
$$
where $\omega$ is an arbitrary real function
chosen so that $\vec{a}$ is regular [3].
Inserting this into the 
Gauss law,  we end up with
the ``Liouville-type'' equation
$$
\bigtriangleup\log\varrho=\pm\alpha\big(\rho-1\big),
\qquad
\alpha={\gamma\over\mu}.
$$

Now, if we want a ``confining'' (stable) and lower-bounded scalar 
potential, $\lambda$ has to be positive. 
Then we see from eq. (2.3) that for 
the upper sign this means $0<\alpha<2$,
 whereas for the lower
sign $-2<\alpha<0$.
In any of the two cases ($\alpha$ positive or negative), the 
coefficient of $(\rho-1)$ in the r. h. s.
 is always positive: in the upper sign, it is
$\alpha$ with $\alpha>0$, in the lower sign,
it is $-\alpha$ with $\alpha<0$. We consider henceforth the 
equation
$$
\bigtriangleup\log\varrho=\vert\alpha\vert\big(\rho-1\big),
\equation
$$

Note that the electric field, $\vec{E}=\vec{\nabla}a_{t}$,
only vanishes for $\mu=\pm\gamma$, i.e., when $\lambda=1$,
which is Manton's case.

Before analyzing the solutions of Eq. (2.6), let us  discuss 
the finite-energy 
conditions.  As it will be derived 
 in the next Section, in the frame where $\vec{J}^{T}=0$,
the energy associated to the Lagrangian (1.1) is 
$$
H=\int\Big\{
\2\big\vert\vec{D}\phi\big\vert^{2}+\2B^{2}
+U(\phi)\Big\}
\,d^{2}\vec{x},
\qquad
U(\phi)={\lambda\over8}\big(1-\vert\phi\vert^{2}\big)^{2}.
\equation
$$
Eliminating the magnetic term $B^2/2$ 
by using the Gauss law (1.4)
results in shifting merely the coefficient of the non-linear 
term,
$$
H=\int\Big\{
\2\big\vert\vec{D}\phi\big\vert^{2}
+
{\Lambda\over8}\big(1-\vert\phi\vert^{2}\big)^{2}\Big\}
\,d^{2}\vec{x},
\qquad
\Lambda=\lambda+{\gamma^{2}\over\mu^{2}}.
\equation
$$

Finite energy requires, just like in the 
Landau-Ginzburg case,
$
\vec{D}\phi\to0
$
and
$
\vert\phi\vert^2\to1
$
so that our objects represent {\it topological vortices}:
The first of these equations implies that the angular component 
of vector potential
behaves asymptotically as $n/r$.
The integer $n$ here is also the
winding number of the mapping defined by the asymptotic values
of $\phi$ into the unit circle,
$$
n={1\over2\pi}\oint_{S}\vec{a}\cdot d\vec{\ell}
={1\over2\pi}\int B\,d^2\vec{x},
\equation
$$
so that the magnetic flux is necessarily quantized, and is related to
the  particle number 
$$
N\equiv\int\big(1-\vert\phi\vert^2\big)\,d^2\vec{x}
={2\mu\over\gamma}\int B\,d^2\vec{x}
=4\pi\big({\mu\over\gamma}\big)n.
\equation
$$
$N$ is conserved since the supercurrent satisfies the 
continuity equation
$
\partial_t\varrho+\vec\nabla\cdot\vec{J}=0
$.

Not surprisingly, our self-duality equations (2.2) can also be obtained
by studying the energy, (2.7).
Using the identity 
$
\big|\vec{D}\phi\big|^2=
\big|(D_1\pm iD_2)\phi\big|^2
\pm B|\phi|^{2}
\pm\vec{\nabla}\times\vec{J}
$
and assuming that the fields vanish at infinity, the integral of
the current-term can be dropped, so that $H$ becomes 
$$
\int\bigg\{
{1\over2}\Big|(D_1\pm iD_2)\phi\Big|^2
+\Big[
\big(\mp{\gamma\over4\mu}+
{\Lambda\over8}\big)(1-|\phi|^2)^2\Big]
\bigg\}d^2\vec{x}
\pm\underbrace{\2\int B\, d^2\vec{x}}_{\pi n},
\equation
$$
which shows that the energy is positive definite  when the square 
bracket vanishes, i.e., 
 for the chosen potential with $\lambda$ as in Eq. (2.3).
In this case, the energy admits a lower ``Bogomolny'' bound,
$H\geq\pi|n|$, with the equality only attained when the
SD equations hold.

Eqn. (2.6) is similar to that of Bogomolny
in the Landau-Ginzburg theory [5] to which it reduces
when $\vert\alpha\vert=1$. 
The proofs of  Weinberg, and of Taubes [6],
carry over literally to show,  for each $n$,
the existence of a $2n$-parameter family of
solutions. 
Radial solutions were studied numerically by
Barashenkov and Harin [7]. 
The solutions behave as in the Bogomolny case.
Write  $\phi=f(r)e^{in\theta}$.
Linearizing the SD Eqns. (2.2), for $\varphi=1-f$  we get
$$
\varphi''+{1\over r}\varphi'-\vert\alpha\vert\varphi=0,
\equation
$$
which is Bessel's equation of order zero. The
solution and its asymptotic behaviour are therefore
$$
\varphi(r)\quad\sim\quad
\matrix{
K_{0}(mr)\hfill
&\sim&
\displaystyle{C\over\sqrt{r}}e^{-mr},
\qquad\hfill
&m=\sqrt{\vert\alpha\vert}.\hfill
\cr}
\equation
$$

The magnetic and electric fields behave in turn as
$$
B={\alpha\over2}(1-f^2)\sim{\alpha}{D\over\sqrt{r}}
e^{-mr},
\qquad
\vec{E}=-\smallover1/{4\mu}\big(\mp1+\alpha\big)\,
\vec\nabla f^2\sim
{G\over\sqrt{r}}e^{-mr}.
\equation
$$

\goodbreak
\chapter{Symmetries and conserved quantities}

Let us remember the definition: an infinitesimal
transformation, represented by a vector field $X^\mu$ on space-time,
is a symmetry when it changes
the Lagrangian by a surface term,
$$
\L\to\L+\partial_{\alpha}K^\alpha
\equation
$$
for some function $K$. 
To each such transformation, N{\oe}ther's theorem
associates a conserved quantity, namely
$$
C=\int\left({\delta\L\over\delta(\partial_{t}\chi)}
\delta\chi-K^t\right)d^2\vec{x},
\equation
$$
where $\chi$ denotes collectively all fields in the theory [8].

Our clue for understanding 
the symmetry properties of the Manton system
is to observe that setting
$$
B^{ext}\equiv
{\gamma\over2\mu},
\qquad
E^{ext}_{k}=-{\epsilon_{kl}{J^{T}}_{l}\over2\mu},
\equation
$$
the equations of motion (1.2-4) become
$$\eqalign{
&i\gamma D_t\phi=
-{1\over2}\vec{D}^2\phi
-{\lambda\over4}\big(1-|\phi|^2\big)\phi,
\ccr
&\epsilon_{ij}\partial_{j}\widetilde{B}=
J_{i}+2\mu\,\epsilon_{ij}\,\widetilde{E}_j,
\ccr
&2\mu\widetilde{B}=-\gamma|\phi|^2,
\cr}
\equation
$$
where 
$
\widetilde{B}=B-B^{ext}
$,
$
\widetilde{E}_{i}=E_{i}-E^{ext}_{i},
$
and
$
D_{\alpha}=\partial_{\alpha}-ia_{\alpha}, 
$
where
$a_{\alpha}=\tilde{A}_{\alpha}+A_{\alpha}^{ext}
$, 
so that 
$\widetilde{F}_{\alpha\beta}=
\partial_{\alpha}\widetilde{A}_{\beta}-
\partial_{\beta}\widetilde{A}_{\alpha}.$
These equations describe
a non-relativistic scalar field
with Maxwell-Chern-Simons dynamics and an external, constant
electromagnetic field [9], [10].

The equivalence of the two models can also be checked
on the respective Lagrangians. That of the external-field 
problem is in fact
$$
\eqalign{
{\cal L}^{ext}
=
&-{1\over2}B^2+\gamma{i\over2}
\big(\phi^*D_t\phi-\phi(D_t\phi)^*\big)
-{1\over2}\big|\vec{D}\phi\big|^2
-U(|\phi|)
\cr
&+\mu\,\big(\tilde{B}\tilde{A}_t+\tilde{E}_2\tilde{A}_1
-\tilde{E}_1\tilde{A}_2\big).
\cr}
\equation
$$
Inserting here $A_{k}^{ext}=-\gamma\epsilon_{kl}x^l/4\mu$
and $A_{t}^{ext}=0$, one gets up to surface terms
the Manton Lagrangian (1.1),
without the transport current term. This latter is finally 
recovered when applying a galilean boost,
$$
\phi(\vec{x}, t)\to e^{-i(\vec{J}^T\cdot\vec{x}
+\2\vert\vec{J}^T\vert^{2}t/\gamma)}
\,\phi\big(\vec{x}+\vec{J}^Tt/\gamma,t\big).
\equation
$$
The necessity of adding a transport current 
 corresponds hence to the 
arisal of an electric field
under a boost.

Before studying the symmetries of the Manton system,
let us recall 
that Jackiw and Pi have shown in Ref. [3] that a pure 
Chern-Simons-matter system with
the non - symmetry - breaking potential
$U=-(g/2)\vert\phi\vert^{4}$
admits the Schr\"odinger group
as symmetry.
The  ``geometric'' action on space-time of this latter
is generated
by the 8-parameter vectorfield
$$ 
\pmatrix{X^t\ccr\vec{X}\cr}
=
\pmatrix{-\chi t^2-\delta\,t-\epsilon\hfill\ccr
\Omega(\vec{x}) 
-\left(\2\delta+\chi t\right)\vec{x}
+t\vec{\beta}+\vec{\gamma}\hfill\cr
},
\equation 
$$
where $\Omega\in{\rm so}(2),\,
\vec{\beta},\vec{\gamma}\in\IR^2,\,
\epsilon,\chi,\delta\in\IR$,
interpreted as rotation, boost, space translation, time translation,
expansion, dilatation.

When this system is put into an external 
field, only those symmetries remain
which are symmetries for this latter in the sense of 
Ref. [8]:
$$
X^\alpha F_{\alpha\beta}^{ext}=\partial_{\beta}\Psi.
\equation
$$

For a constant 
electric and magnetic field, it is readily seen that
the Schr\"odinger symmetry, {\it acting as in (3.7) on spacetime}, 
is  broken.
Only the time and space translations survive in general:
Eqn. (3.8) is satisfied with
$$\matrix{
\Psi=\vec{x}\cdot\vec{E}^{ext}\epsilon
\hfill
&\hbox{for time translations},\hfill
\ccr
\Psi=B^{ext}\vec{\gamma}\times\vec{x}+t\vec{\gamma}\cdot\vec{E}^{ext}
\qquad
\hfill
&\hbox{for space translations}.\hfill
\cr}
\equation
$$

Exceptions may also occur, namely,

$\bullet$ When $B^{ext}=0$, we also have boosts;

$\bullet$ When $\vec{E}^{ext}=0$, we also have rotations.

In what follows, we only consider the case $B^{ext}\neq0$.

Now the pure Chern-Simons  system in a constant external 
electromagnetic field admits another, ``hidden''
symmetry [9], [10], [11], with generators
$$
\matrix{
\cos\omega t
\pmatrix{0
\ccr
\cos\omega t\,\gamma_{1}-\sin\omega t\,\gamma_{2}
\ccr
\sin\omega t\,\gamma_{1}+\cos\omega t\,\gamma_{2}
\cr}\qquad\hfill
&\gamma_1,\ \gamma_2\in\IR\hfill
&\hbox{``translations''},\hfill
\cr\cr\cr
\displaystyle-{\sin\omega t\over\omega}\pmatrix{0
\ccr
\cos\omega t\,\beta_1-\sin\omega t\,\beta_{2}
\ccr
\sin\omega t\,\beta_{1}+\cos\omega t\,\beta_{2}
\cr}
\hfill
&\beta_1,\ \beta_2\in\IR\hfill
&\hbox{``boosts''},\hfill
\cr\cr\cr
\qquad\pmatrix{0\ccr
-\Omega\big(x_{2}
-(E_1^{ext}/B^{ext})t\big)
\ccr
\Omega\big(x_{1}
+(E_2^{ext}/B^{ext})t\big)
\cr}
\hfill
&\Omega\in\IR\hfill
&\hbox{``rotations''},\hfill
\cr}
\equation
$$
(where $\omega=\2B^{ext}$),
as well 3 more generators, we call ``dilatations'',
``expansions'' and ``time translations''.
(Their rather complicated  expressions 
are here omitted, since they are
 not needed for our purposes).
Surprisingly, this algebra turns out to be abstractly
isomorphic to the Schr\"odinger algebra, 
as anticipated by the terminology.
The action of this ``hidden'' Schr\"odinger algebra
on spacetime is different from the geometric one in (3.7), though.
The ``hidden rotations'' look, e.g., rather as cycloidal motions;
they reduce to 
the ``ordinary'' (``geometric'') rotations 
only when $\vec{E}^{ext}=0$.  
Note also that the geometric translations in 
(3.7) are related to the ``hidden'' ones in (3.10)
according to
$$\eqalign{
&\hbox{(geometric translation)}_{i}
=
\cr
&\hbox{(``hidden translation'')}_{i}
+\omega\epsilon_{ij}\,
\hbox{(``hidden boost'')}_{j}.
\cr}
\equation
$$
\goodbreak

Let us now return to the Manton system.
The ``non-relativistic Maxwell term'' $B^{2}$ is 
Schr\"odinger invariant.
The potential $(\lambda/4)\big(1-\vert\phi\vert^2\big)^2$
breaks, however, the ``hidden''
dilatations, expansions and even time translations
(that's why we did not write them at all).
We are hence left with a five-parameter subgroup of
the ``hidden'' Galilei group made of (``hidden'')
``translations'', ``boosts'' and ``rotations'',
which acts as symmetry for the Manton system.
(Note that the new, ``geometric'' time translations do
{\it not} belong to this unbroken subgroup).

Having determined the symmetries of our problem,
we now turn to the associated conserved quantities.
The formula (3.2) can also be written in terms of the
energy-momentum tensor $T_{\alpha\beta}$  as
$$
C_{X}=\int T_{0\alpha}X^\alpha d^{2}\vec{x}.
\equation
$$

The energy-momentum tensor $T_{\alpha\beta}$
 is readily derived by modifying the
expression found by Jackiw and Pi in the pure Chern-Simons case [3].
Returning to the original variables,
we find:
$$\eqalign{
&T_{00}=
\2\big\vert\vec{D}\phi\big\vert^{2}
+
{\Lambda\over8}\big(1-\vert\phi\vert^{2}\big)^{2}
+
\vec{J}^T\cdot\vec{a}
-\2\vert\vec{J}^T\vert^2\vert\phi\vert^2,
\qquad
\Lambda=\lambda+\big({\gamma\over\mu}\big)^2,
\ccr
&T_{k0}=-\2\big[(D_{t}\phi)^*D_{k}\phi+(D_{t}\phi)(D_{k}\phi)^*\big]
+\epsilon_{kj}E_{j}B
-a_{0}J^{T}_{k}
-\smallover1/{2\gamma}\vert\vec{J}^T\vert^2J_{k},
\ccr
&T_{0k}=-\gamma\smallover{i}/{2}\big(\phi^{*}D_{k}\phi
-\phi(D_{k}\phi)^{*}\big)
+\gamma a_{k}-\gamma J_{k}^T\vert\phi\vert^{2},
\ccr
&T_{ij}=\2\Big((D_{i}\phi)^{*}D_{j}\phi+(D_{j}\phi)^{*}D_{i}\phi
-\delta_{ij}\big\vert\vec{D}\phi\big\vert^{2}\Big)+J^T_{i}a_{j}
- J_{i}J_{j}^T,
\ccr
&\qquad
+\smallover1/4\big(
\delta_{ij}\bigtriangleup-2\partial_{i}\partial_{j}\big)\varrho
+\delta_{ij}\left[T_{00}-\gamma a_{0}-2\vec{a}\cdot\vec{J}^T
+{\lambda\over4}\big(\vert\phi\vert^{2}-1\big)
+\2\vert\vec{J}^T\vert^2\vert\phi\vert^2\right].
\ccr}
\equation
$$
(The energy-momentum tensor has been improved so that
the integrals below converge). It
satisfies the continuity equation
$
\partial_{t}T_{0\mu}+\partial_{k}T_{k\mu}=0.
$
Note that $T_{0j}\neq T_{j0}$ and that  $T_{ij}$ is only symmetric 
in the frame where $\vec{J}^T=0$; this is obviously related to 
the breaking of ``ordinary'' rotational symmetry.

Note that, unlike in a relativistic theory,
the energy-momentum tensor is not traceless
but satisfies instead
$$
\sum_{i}T_{ii}=
2T_{00}+{\lambda\over2}\big(\vert\phi\vert^2-1\big)
-2\gamma a_{0}
-3\vec{J}^T\cdot\vec{a}-\vec{J}\cdot\vec{J}^T
+\vert\vec{J}^T\vert^2\vert\phi\vert^2.
\equation
$$
(This is consistent with the breaking of the
Schr\"odinger symmetry).

For the surviving geometric symmetries
we find the conserved energy and momentum,
$$\matrix{
H=\displaystyle
\int\left\{
\2\big\vert\vec{D}\phi\big\vert^{2}
-
\2\big\vert\vec{J}^T\big\vert^2\vert\phi\vert^2
+
{\Lambda\over8}\big(1-\vert\phi\vert^{2}\big)^{2}
-
\big(\vec{x}\times\vec{J}^T\big)B
\right\}d^2x,\hfill&
\ccr
\qquad\quad\Lambda=\lambda+\big({\gamma\over\mu}\big)^2,\quad
\hfill&
\hbox{energy}\hfill
\ccr
{\cal P}_{i}=
\gamma\displaystyle\int\Big\{
J_{i}-J^T_{i}\vert\phi\vert^{2}
+\epsilon_{ij}\big({x}^j-t{J^T_{j}\over\gamma}\big)B\Big\}\,
d^{2}\vec{x}\hfill
&\hbox{momentum}\hfill
\cr}
\equation
$$

The energy integral converges, since
$\vert\vec{D}\phi\vert\to\vert\vec{J}^T\vert$
and $\vert\phi\vert^2\to1$ when $\vert\vec{x}\vert\to\infty$.
Note also the extra piece
proportional to the magnetic field $B$ in the momentum.
Because of this piece, the Dirac
bracket of the momenta satisfies
$$
\big\{{\cal P}_{1}, {\cal P}_{2}\big\}=
\gamma\int\!Bd^2x
=\gamma\,2\pi\, n,
\equation
$$
rather then vanishes  (see the Appendix). 
(This has been found also by
Barashenkov and Harin [7] in their model).

The conserved quantities associated 
to the the unbroken part of the ``hidden'' symmetry (3.10)
 can also be readily calculated using (3.12).
The explicit expressions are not illuminating 
and therefore omitted.   
It is, however, interesting to point out the
relation between the ``geometric'' momentum, $\vec{\cal P}$,
the ``hidden momentum'' $\vec{p}$, and ``hidden boost''
$\vec{g}$,
$$
{\cal P}_{i}=p_{i}+\omega\epsilon_{ij}g_{j},
\qquad
\omega\equiv\2B^{ext},
\equation
$$
which is plainly the analog of the relation (3.11) between the 
generating  vectorfields. This explains 
the unusual commutation relations (3.16).
Our hidden \lq\lq translations''
and \lq\lq boosts'' satisfy in fact
$$
\big\{p_{i}, p_{j}\big\}
=0=
\big\{g_{i}, g_{j}\big\},
\qquad
\big\{ g_{j}, p_{i}\big\}=\gamma\,N\delta_{ij},
\equation
$$
where $N$ is the particle number (2.10).

For the ``hidden angular momentum'', we get in turn
$$\matrix{
M=\gamma\displaystyle\int\Big\{\vec{x}\times\big(\vec{J}-\vec{J}^T\vert\phi\vert^{2}\big)
-\2r^2B\hfill&
\ccr
\quad-{t\over\gamma}\left[\vec{J}^{T}\times\vec{J}
-(\vec{x}\cdot\vec{J}^T)B\right]
-\2\big({t\over\gamma}\big)^2\vert\vec{J}^T\vert^{2}B\Big\}
\,d^{2}\vec{x}\hfill
\hfill
&\hbox{``angular momentum''}.
\cr}
\equation
$$

Note here the extra piece
proportional to the total magnetic field $B$. Note also that
the integrals converge since $\vec{J}\to\vec{J}^T$, and 
$\vert\phi\vert\to1$ at spatial infinity.
For $\vec{J}^T=0$, (3.19) reduces to
 the well-known formul{\ae}
in a magnetic field.

\vfill\eject

\goodbreak
\chapter{Spinor vortices}

In  Ref. [12], we found non-relativistic, 
spinor vortices in pure Chern-Simons theory. 
Below we generalize our construction to the magnetic type
non-relativistic Maxwell-Chern-Simons theory of Manton's type.
Let $\Phi$ denote a 2-component Pauli spinor.
We posit the following equations of motion.
$$\left\{
\matrix{
i\gamma D_t\Phi=-{1\over2}\big[\vec{D}^2+B\sigma_3\big]\Phi\hfill
&\hbox{Pauli eqn.}\hfill
\ccr
\epsilon_{ij}\partial_{j}B
=
J_{i}-J^{T}_{i}+2\mu\,\epsilon_{ij}\,E_j\qquad\quad\hfill
&\hbox{Amp\`ere's eqn.}\hfill
\ccr
2\mu B=\gamma\big(1-|\Phi|^2\big)\hfill
&\hbox{Gauss' law}\hfill
\cr}\right.
\equation
$$
where the current is now 
$$
\vec{J}={1\over2i}\Big(
\Phi^\dagger\vec{D}\Phi-(\vec{D}\Phi)^\dagger\Phi\Big)
+\vec{\nabla}\times\Big({1\over2}\,\Phi^\dagger\sigma_3\Phi\Big).
\equation
$$
The system is plainly non-relativistic, 
and it admits self-dual vortex solutions, 
as we show now.
The transport current can again be eliminated by a
galilean boost. For fields which are static in
the frame where $\vec{J}^T=0$, the equations of motion become
$$\left\{\eqalign{
&\big[{\2}(\vec{D}^2+B\sigma_3)+\gamma a_t\big]\Phi=0,
\ccr
&\vec\nabla\times B=
\vec{J}+2\mu\,\vec\nabla\times a_t,
\ccr
&2\,{\mu\over\gamma}B=1-\Phi^\dagger\Phi.
\cr}\right.
\equation
$$

Let us now attempt to solve the static equations (4.3)
by the first-order Ansatz
$$
\big(D_1\pm iD_2\big)\Phi=0.
\equation
$$ 
Then
$$
\vec{D}^2=\mp B
\and
\vec{J}
=
{\2}\vec\nabla\times\Big[\Phi^\dagger(\mp1+\sigma_3)\Phi\Big],
\equation
$$
so that the static Pauli equation requires
$$
\Big[(\mp 1+\sigma_3)B+2\gamma a_t\Big]\Phi=0.
\equation
$$

Let us decompose $\Phi$ into chiral components,
$$
\Phi=\Phi_{+}+\Phi_-,
\where
\Phi_+=\pmatrix{0\cr\chi\cr}
\and
\Phi_-=\pmatrix{\varphi\cr0\cr}.
\equation
$$
Eqn. (4.6) requires that $\Phi$ has a definite chirality.
One possibility would be $\Phi_+=0$
for the upper sign and $\Phi_-=0$ for the lower sign, as in 
Ref. [12].
In both cases,  $a_t$ would have to vanish. It is, however,
easely seen to be inconsistent with Amp\`ere's law.

Curiously, there is another possibility: one can have
$$
a_t=\pm\smallover1/\gamma\,B
\and
\matrix{
\Phi_-=0\quad\hfill&\hbox{i.e.}\;\Phi\equiv\Phi_+\quad\hfill
&\hbox{for the upper sign}\hfill
\ccr
\Phi_+=0\quad&\hbox{i.e.}\;\Phi\equiv\Phi_-\quad\hfill
&\hbox{for the lower sign}\hfill
\cr}.
\equation
$$
Then
$
\vec{J}
=
\mp\vec\nabla\times\big|\Phi_{\pm}\big|^2,
$
so that Amp\`ere's law requires
$$
\vec{\nabla}\times\Big(\big[1\mp\smallover{2\mu}/{\gamma}\big]B
\pm\big|\Phi_{\pm}\big|^2\Big)=0.
\equation
$$
But now $\big|\Phi_{\pm}\big|^2=\big|\Phi\big|^2$, which is 
$1-(2\mu/\gamma)B$
by the Gauss law, so that (4.9) holds when
$$
\alpha\equiv\pm{\gamma\over\mu}=4.
\equation
$$
In conclusion, for the particular value
(4.10), the second-order field equations
can be solved by solving one or the
other of the first-order equations in (4.4).
Now these latter conditions fix the gauge potential as
$$
\vec{a}=\pm\2\vec{\nabla}\times\log\varrho,
\qquad
\varrho\equiv\big|\Phi\big|^2
=\big|\Phi_{\pm}\big|^2.  
\equation
$$
Then the Gauss law yields
$$
\bigtriangleup\log\varrho=4(\varrho-1),
\equation
$$
which is again the ``Liouville-type'' equation (2.6) 
we studied before. Note that the sign 
--- the same for both choices ---
is automatically positive and equal to $4$.

The equations of motion (4.1) can be derived from the Lagrangian
$$
\eqalign{
{\cal L}=
&-{1\over2}B^2+{i\gamma\over2}
\big[\Phi^\dagger(D_t\Phi)-(D_t\Phi)^\dagger\Phi\big]
-{1\over2}(\vec{D}\Phi)^\dagger(\vec{D}\Phi)
\cr
&+{B\over2}\Phi^\dagger\sigma_3\Phi
+\mu\big(Ba_t+E_2a_1-E_1a_2\big)
-\gamma a_t-\vec{a}\cdot\vec{J}^{T}.
\cr}
\equation
$$
Then, in the frame where $\vec{J}^T=0$, 
the associated energy is
$$
H=
{1\over2}\int\left\{
B^2
+\big|\vec{D}\Phi\big|^2
-B\,\Phi^\dagger\sigma_3\Phi
\right\}\,d^2\vec{x}.
\equation
$$
Using the identity
$$
\big|\vec{D}\Phi\big|^2=\big|(D_{1}\pm iD_{2})\Phi\big|^2
\pm B\,\Phi^\dagger\Phi
\equation
$$
(up to surface terms), the energy is rewritten as
$$
H=
\2\,\int\left\{B^2
+\big|(D_{1}\pm iD_{2})\Phi\big|^2
-B\Big[\Phi^\dagger(\mp1+\sigma_3)\Phi\Big]
\right\}\,d^2\vec{x}.
$$
Eliminating $B$ using the Gauss law, we get finally,
for purely chiral fields, $\Phi=\Phi_{\pm}$, 
$$
H=
\2\,\int\left\{
\big|(D_{1}\pm iD_{2})\Phi_{\pm}\big|^2
+{\gamma\over4\mu}\big[\mp4+{\gamma\over\mu}\big]
\big(1-\vert\Phi_{\pm}\vert^2\big)^{2}
\right\}\,d^2\vec{x}
\pm\,\int B\,d^2\vec{x}.
\equation
$$
Here the last integral yields the topological charge
$\pm2\pi n$.
The integral is positive definite when $\pm\gamma/\mu\geq4$
depending on the chosen sign, yielding the Bogomolny bound
$H\geq2\pi\vert n\vert$. 
The Pauli term results hence in {\it doubling} 
the Bogomolny bound with respect to the scalar case. 
The bound can be saturated when $\pm\gamma/\mu=4$
and the self-dual equations (4.4) hold.

\goodbreak
\chapter{Relativistic models and their non-relativistic limit}

In relativistic Maxwell-Chern-Simons theory
self-dual solutions only arise  when an auxiliary neutral field 
$N$ is added [13]. Here we present a
  model of the type considered by Lee, Lee and Min,  which
(i) is relativistic; (ii) can be made self-dual;
(iii) its non-relativistic limit is the Manton model
presented in this paper.
Let us consider in fact $(1+2)$-dimensional Minkowski space
with the metric  $(c^{2}/\gamma,-1,-1)$ where $\gamma>0$ is a constant.
Let us chose the Lagrangian
$$
\L_{R}=-\smallover1/4F_{\mu\nu}F^{\mu\nu}
+
\smallover\mu/2\epsilon^{\mu\nu\rho}F_{\mu\nu}a_\rho
+
\big(D_\mu\psi\big)\big(D^\mu\psi\big)^*
+
a^{\mu}{J^T}_{\mu}
+
\smallover{\gamma}/{2c^{2}}\partial_\mu N\partial^\mu N
-V.
\equation
$$

Here $N$ is  an auxiliary neutral field, which we chose real. 
We have also included the term $a^{\mu}{J^T}_{\mu}$
where the Lorentz vector ${J^T}_{\mu}$ represents the relativistic 
generalization of Manton's transport current. We chose ${J^T}_{\mu}$
to be time-like, $\I^{2}\equiv{\gamma\over c^2}{J^T}_{\mu}{J^T}^{\ \mu}>0$.
 Our choice for the potential is
$$
V={\beta\over2}\big(\vert\psi\vert^{2}-2\vert\mu\vert N
-{\I\over2m\gamma}\big)^2
+
{\gamma\over c^2}\big(N+mc^2\big)^2\vert\psi\vert^2
-
(N+mc^2){\rm I},
\equation
$$
where $\beta>0$. 
Although the potential is {\it not} positive definite,
this will cause no problem when the Gauss law is
taken into account, as it  will be explained later.
Note that a similar behaviour has already been encountered before [14].
This Lagrangian is clearly Lorentz-invariant so that the model is 
indeed relativistic.

The associated equations of motion read
$$
\matrix{
D_\mu D^\mu\psi+\displaystyle{\partial V\over\partial\psi^*}=0,
\hfill
&\hbox{Non-linear Klein-Gordon eqn.}\hfill
\ccr
{\gamma\over c^2}\partial_0F_{0i}+\epsilon_{ij}\partial_j F_{12}
+2\mu\epsilon_{ij}F_{0j}-J_i+{J^T}_{i}=0,\quad
\hfill
&\hbox{Amp\`ere's law}\hfill
\cr\cr
{\gamma\over c^2}\partial_i F_{0i}+2\mu F_{12}
=
{\gamma\over c^2}\big(J_0-{J^T}_{0}\big),
\hfill
&\hbox{Gauss' law}\hfill
\cr\cr
{\gamma\over2c^2}\partial_\mu\partial^\mu N+
\displaystyle{\partial V\over\partial N}=0
\hfill
&\hbox{auxiliary eqn. for\ } $N$.\hfill
\ccr}
\equation
$$

One can always choose a Lorentz frame where the spatial
components of the
transport current vanishes, ${J^T}_{\mu}=(-{c^2\over\gamma}\I,0)$.
In such a frame, using the Gauss law, for the energy we find
$$\eqalign{
&H_{R}=
\ccr
&\int\left\{\smallover\gamma/{2c^2}\,\vec{E}\strut^2+\2B^2
+\smallover\gamma/{c^2}\,\big|D_0\psi\big|^2+\big|\vec{D}\psi\big|^2
+\smallover{\gamma^2}/{2c^4}\,\big(\partial_0N\big)^2
+\smallover\gamma/{2c^2}\,\big(\vec\nabla N\big)^2
+V\right\}d^2x,
\cr}
\equation
$$
where we used the obvious notations
$E_i=F_{0i}$, $B=F_{12}$ and we have assumed that the surface terms,
$$
\smallover\gamma/{c^2}\,\vec\nabla\cdot\big(a_{0}\vec{E}\big)
+\mu\vec{\nabla}\times\big(a_{0}\vec{a}\big),
\equation
$$
fall off sufficiently rapidly at infinity.
To get finite energy, we require that the energy density go
to zero at infinity. Note that 
$
\vert D_{0}\psi\vert^2
$
does {\it not} go to zero at infinity, because
$J_{0}=(-i)\big(D_{0}\psi\psi^*-\psi(D_{0}\psi)^*\big)$
has to go to $J_{0}^T\neq0$ at spatial infinity.
This term
combines rather with the last two
terms in the potential.
At spatial infinity, the energy density becomes
the sum of positive terms.
Requiring that all these terms go to zero allows us to conclude that
finite energy requires 
$$
\vert\vec{E}\vert\to0,
\qquad
B\to0,
\qquad
\vert\psi\vert^{2}\to{\I\over2m\gamma},
\qquad
N\to0.
\equation
$$

Using the Bogomolny trick and the Gauss' law as in Eqn. (5.3),
the term linear in $N$ in the potential gets absorbed.
Then the energy is re-written, for the particular value $\beta=1$, as
$$
\eqalign{
H_{R}=\int&
\left\{
\smallover\gamma/{2c^2}\,\big[\vec{E}+\vec\nabla N\big]^2
+\2\big[B+\epsilon(|\psi|^2-2\vert\mu\vert N-{\I\over2m\gamma})\big]^2
\right.
\ccr
&\left.
+
\smallover\gamma/{c^2}\big|D_0\psi+i(N+mc^{2})\psi\big|^2
+\big|(D_1+i\epsilon D_2)\psi\big|^2
+
\smallover\gamma^{2}/{2c^4}\big[\partial_0N\big]^2
\right\}d^2x
\ccr
&\quad
-\epsilon\big(2\vert\mu\vert mc^2-{\I\over2m\gamma}\big)
\underbrace{\int B\,d^2x}_{\hbox{\small flux}},
\cr}
\equation
$$
where $\epsilon$ is the sign of $\mu$.
The last term is topologic, labelled by the winding number,
$n$, of $\psi$. 
Due to the presence of $c^2$, it seems to be reasonable to assume that 
the coefficient in front of the magnetic flux is positive.
Then, chosing $n<0$ for $\epsilon\equiv$sign$(\mu)>0$ and
$n>0$ for $\epsilon\equiv$ sign$(\mu)<0$ respectively,
the energy admits hence the ``Bogomolny'' bound
$$
H_R\geq
\big(2\vert\mu\vert mc^2-{\I\over2m\gamma}\big)\,2\pi \vert n\vert.
\equation
$$

The absolute minimum  is attained by those configurations which
solve the ``Bogomolny'' equations
$$
\eqalign{
&\partial_{0}N=0,
\ccr
&\vec\nabla N+\vec{E}=0,
\ccr
&
D_{0}\psi+i(N+mc^{2})\psi=0,
\ccr
&
\big(D_{1}+i\epsilon D_{2}\big)\psi=0,
\cr
&
B=\epsilon\big({\I\over2m\gamma}-\vert\psi\vert^2+2\vert\mu\vert 
N\big).
\cr}
\equation
$$
It can also be checked directly that the solutions of the 
self-duality equations (5.9) solve the second-order field equations 
(5.3), when the  gauge fields are static and the matter field is
of the form 
$$
\psi=e^{-imc^2t}\times\hbox{(static)}, 
$$
cf. Ref. [3].
These equations are similar to those of by Lee et al., and
could be studied numerically as in Ref. [13].
Note that, just like in the case studied by Donatis and Iengo [15],
the solutions are {\it chiral} in that the winding number and the sign 
of $\mu$ are correlated.

Let us stress that for getting a non-zero electrical field, the 
presence of a non-vanishing auxiliary field $N$ is essential.
For $N=0$ we get rather a  self-dual extension of the model of
 Paul and Khare [16], whose vortex solutions
are purely magnetic.

Now we show that the non-relativistic limit 
of our relativistic model presented above is
precisely the Manton model. To see this, let us set
$$
\psi={1\over\sqrt{2m}}e^{-imc^{2}t}\,\phi.
\equation
$$
The transport current is the long-distance limit of the supercurrent,
 ${J^T}_\mu=\lim_{r\to\infty}J_\mu$. But
$
\lim_{c\to\infty}J_0/c^2=-\vert\phi\vert^{2},
$
so we have
$$
\lim_{c\to\infty}{J^T}_0/c^2=
-\lim_{r\to\infty}\vert\phi\vert^{2}=
-\lim_{c\to\infty}{\I\over\gamma}\equiv-\alpha.
\equation
$$
Then the standard procedure (described, e. g., in [3]), 
yields, after dropping the term $mc^2\I$, the 
non-relativistic expression
$$
\eqalign{
{\cal L}_{NR}=
&-{1\over2}B^2+\gamma{i\over2}\big(\phi^*D_t\phi-\phi(D_t\phi)^*\big)
-{1\over2m}\big|\vec{D}\phi\big|^2
\cr
&+\mu\big(Ba_t+E_2a_1-E_1a_2\big)
-\gamma a_t-\vec{a}\cdot\vec{J}^{T}
\cr
&-\Big\{{\beta\over8m}\big(\alpha-|\phi|^2+
4m\vert\mu\vert N\big)^2
-\gamma\big(\alpha-|\phi|^2\big)N\Big\}.
\cr}
\equation
$$

Note that there is no kinetic term left for the auxiliary field $N$. 
It can therefore be eliminated altogether
by using its equation of motion,
$$
4\mu^2\beta N=\big(\gamma-{\vert\mu\vert\beta\over m}\big)
\big(\alpha-\vert\phi\vert^2\big).
\equation
$$
Inserting this into the potential, this latter becomes
$$
\big({\gamma\over4\vert\mu\vert 
m}-{\gamma^2\over8\mu^2\beta}\big)
\big(\alpha-\vert\phi\vert^2\big)^{2}.
\equation
$$
For $\alpha=1$ and $m=1$ in particular, we get precisely
the Manton Lagrangian (1.1) with
$$
\lambda={2\gamma\over\vert\mu\vert}-{\gamma^{2}\over\mu^{2}\beta}.
\equation
$$

The non-relativistic limit of the equations of movement (5.3)
is (1.2-4), as it should be.

$\bullet$
In Amp\`ere's law, the first term
$(\gamma/c^2)\partial_{0}F_{0i}$ can be dropped;
setting (5.10), the relativistic current  becomes the
non-relativistic expression 
$
\vec{J}=({1/2i})\big(\phi^*\vec{D}\phi-\phi(\vec{D}\phi)^*\big);
$

$\bullet$ In Gauss' law, the first term
$(\gamma/c^2)\partial_{i}F_{0i}$ can be dropped;
the time-component of the currents behave, as already noticed, as
$$
\lim_{c\to\infty}J_0/c^2=-\vert\phi\vert^{2},
\and
\lim_{c\to\infty}{J^T}_0/c^2=
-\alpha=-1.
$$

$\bullet$ In the equation for the auxiliary field $N$ the first term
$(\gamma/c^2)\partial_\mu\partial^\mu N$
can be dropped and the $c\to\infty$ limit of $\p V/\p N=0$
is (5.13);

$\bullet$ Finally, setting (5.10)  
in the nonlinear Klein-Gordon equation 
and using the equation of motions 
(5.13) for $N$, a lengthy but straightforward calculation yields
the non-linear Schr\"odinger equation (1.2), as expected.

Note also that, for the self-dual value 
$\beta=1$ (when $\lambda$ in (5.15) 
becomes (2.3)), 
the non-relativistic limit of the (relativistic) 
self-dual equations (5.9)
fixes $a_0$ and $N$ as
$$
a_{0}=N=
\big(-{\epsilon\over4\mu}+{\gamma\over4\mu^2}\big)
\big(1-\vert\phi\vert^2\big).
\equation
$$
which is consistent with Eq. (2.4). The other equations reduce in 
turn to our non-relativistic self-dual equations (2.2).

\goodbreak
\chapter{Further models}

\kikezd{A}. As already said, some of our formul{\ae} bear a
strong ressemblence to those of
Barashen\-kov and Harin [7], who study 
the system described by
$$
\eqalign{
&{\cal L}=
{1\over2}E^{2}-{1\over2}B^{2}
+{i\over2}\big(\phi^*D_t\phi-\phi(D_t\phi)^*\big)
-{1\over2}\big|\vec{D}\phi\big|^2
-{\lambda\over8}\big(1-|\phi|^2\big)^2
\cr
&+\mu\big(Ba_t+E_2a_1-E_1a_2\big)
-\gamma a_t.
\cr}
\equation
$$

This Lagrangian only differs from the Manton model in that it 
contains the
full Maxwell term, while the transport term $\vec{J}^T\cdot\vec{a}$
is missing.
The Barashen\-kov-Harin model has, therefore, no clear
symmetry: the (full)
Maxwell term is Lorentz invariant;
the matter term is Galilei invariant;
the Chern-Simons term is invariant with respect to any
diffeomorphism.
Finally, their naked $-\gamma\cdot a_{t}$ term breaks both the Lorentz
and Galilei invariance.

The energy of the Barashen\-kov-Harin model is
$$
H=\int\Big\{\2\vec{E}\strut^{2}+\2B^{2}
+\2\big\vert\vec{D}\phi\big\vert^{2}
+{\lambda\over8}\big(1-\vert\phi\vert^{2}\big)^{2}\Big\}
\,d^{2}\vec{x},
\equation
$$
while their Gauss law reads
$$
\vec\nabla\cdot\vec{E}
-2\mu B-|\phi|^2+\gamma=0.
\equation
$$

Now the presence of $\vec{E}^2$ in the energy and
of $\vec{\nabla}\cdot\vec{E}$
in their Gauss' law only allows, 
just like in other relativistic models, 
a vanishing electric field, [17]
--- unless an auxiliary field is added [13], [18].

After putting the electric field to zero by hand, the remaining
Barashen\-kov-Harin equations coincide with ours. 
It is hence precisely the {\it absence} of the electric terms
--- dictated by the requirement of a 
{\it consistently non-relativistic theory} ---
which opens the door for solutions with nonzero
electric field in Manton's model.

Let us note in conclusion that the more general type of self-duality
with two non-vanishing components [19] 
only works in the pure Chern-Simons case, and breaks
down when the Maxwell term is present, due 
to $\vec\nabla\times{B}$ in Amp\`ere's law.

\kikezd{B}. Let us mention that a consistently non-relativistic
Maxwell-Chern-Simons model has also been considered before, 
namely in a ``non-relativistic Kaluza-Klein-type'' framework 
 [20]. There one starts with a
four-dimensional (relativistic) coupled Maxwell-Chern-Simons
theory. When the theory is reduced to 2+1 dimensions
by factoring out a
lightlike, covariantly-constant direction, one gets a 
non-relativistic system 
with  equations of motion
$$
\eqalign{
&iD_t\phi=
-{1\over2}\vec{D}^2\phi
+{\delta U\over\delta\phi^{*}},
\cr
&\epsilon_{ij}\partial_{j}B=
J_{i}+2\mu\,\epsilon_{ij}\,E_j,
\ccr
&2\mu B=-\vert\phi\vert^{2},
\cr}
\equation
$$
Note the absence of the
transport current in Amp\`ere's law and that
the Gauss law has the Jackiw-Pi form. 
This system can be solved along the same lines
as in Manton's case:
Using the Gauss law, the self-duality equations
$$\eqalign{
&(D_{1}\pm iD_2)\phi=0,
\cr
&2\mu B=-|\phi|^2,
\cr}
\equation
$$
are readily seen to solve the field equations,
provided the potential is
$$
U(\phi)=
-{\lambda\over8}\,\vert\phi\vert^4,
\qquad
\lambda\equiv
\smallover1/{\mu^{2}}\mp\smallover2/{\mu}.
\equation
$$
\vskip-2mm\noindent

Then, inserting  
$$
\vec{a}=\pm\2\vec\nabla\times\log\varrho+\vec\nabla\omega
\and
a_{t}=\smallover1/4
\big(\pm\smallover1/\mu-\smallover1/{\mu^{2}}\big)\varrho,
\equation
$$
into the Gauss' law, we get the Liouville equation,
$$
\bigtriangleup\log\varrho=\pm{1\over\mu}\,\varrho.
$$
Regular solutions arise when the r. h. s. is negative.
Hence the upper sign works for $\mu<0$ and
the lower sign works for $\mu>0$.
For both signs, the particle density $\varrho=\vert\psi\vert^2$
satisfies finally
$$
\bigtriangleup\log\varrho=-{1\over\vert\mu\vert}\,\varrho,
\equation
$$
which is precisely the problem
solved by Jackiw and Pi in the pure Chern-Simons case [3].
Note that 
$
\lambda=1/\mu^2\pm2/\mu
$
is always positive
so that the potential (6.6) is attractive.

The same conclusion can be reached by noting that 
the energy of this system is simply
$$
H=\int\Big\{
{1\over2}\big\vert\vec{D}\phi\big\vert^{2}
-{g\over2}\vert\phi\vert^{4}\Big\}
\,d^{2}\vec{x},
\qquad
g={\lambda\over4}-{1\over4\mu^{2}},
\equation
$$
which is again of the Jackiw-Pi form, the only effect of the
Maxwell field being a shift,
$
\lambda
\to
\lambda-{1/\mu^2},
$
in the coefficient of the non-linearity.
The latter model is known however to be self-dual precisely when
the coefficient of the Chern-Simons term and the non-linearity
are related as $g=\mp{1/2\mu}>0$, which yields the
value (6.6) for $\lambda$ once again.

Note that  this system admits 
the full ``geometric'' Schr\"odinger symmetry (3.7),
just like in the Jackiw-Pi case.
The ``conformal'' symmetry is indicated by 
the energy-momen\-tum tensor satisfying now
$\sum_{i}T^{ii}=2T^{00}$, and
then the same argument as in the Jackiw-Pi case shows that all 
static solutions are  necessarily self-dual [3].

Let us point out in conclusion that the Manton model
  is in fact the non-relativistic field 
theoretical generalization
of the static system introduced by Girvin [21] in his 
\lq\lq Landau-Ginzburg'' theory for the Quantum Hall Effect.

\goodbreak
\kikezd{Acknowledgements}.
 M. H. and J.-C. Y. acknowledge
the Laboratoire de Math\'emathiques et de Physique Th\'eorique
of Tours University for hospitality.  
They are also indebted to the French Government
and  the Gouvernement de La C\^ote d'Ivoire
respectively, for doctoral scholarships.
P. H. would like to thank 
Professors C. Duval, J. Balog and P. Forg\'acs
for illuminating discussions.
We are also indebted to the Referee for pointing out
a sign error leading to unphysical solutions
in the first version of our paper.

\vskip5mm
\goodbreak
\vfill\eject
\noindent{\bf Appendix~: Dirac brackets} 

Using the expression (3.3) of the momenta and setting
$\phi=fe^{i\omega}$, the Dirac bracket 
$\big\{{\cal P}_{1}, {\cal P}_{2}\big\}$ becomes
$$
\big\{{\cal P}_{1}, {\cal P}_{2}\big\}
=
\gamma
\int
\vec\nabla\times\big[f^{2}(\vec\nabla\omega-\vec{J}^T)\big]
d^{2}\vec{x}=
\gamma
\oint_{S}\big(\vec\nabla\omega-\vec{J}^T\big)\cdot d\vec{\ell}
\eqno(A.1)
$$
by using Stokes' theorem, where $S$ is the circle at infinity.
Now if $\vec{J}^T=0$ then the current $\vec{J}$ goes to zero
at infinity. $\vec{A}$ is hence a pure gauge and the integral
in A.1 yields ($2\pi$ times) the winding number,
$$
\big\{{\cal P}_{1}, {\cal P}_{2}\big\}
=
2\pi n\, \gamma.
\eqno(A.2)
$$
For $\vec{J}^T\neq0$,
a boost  with velocity $\vec{\beta}=\vec{J}^T/\gamma$
absorbes the transport current into the phase and we are back in the 
previous case. The result (A.2) is hence valid in all cases.

The conserved quantities, denoted by $\vec{p}$ and
$\vec{g}$, associated to \lq\lq hidden boosts and translations''
are readily found by inserting (3.10) into (3.12). Then a lengthy but
elementary calculation yields the commutation relations (3.18).
For example,
$$\eqalign{
\big\{g_1,p_1\big\}=\gamma\int
&\Big(
\cos^2\omega t\, x_1\partial_1(|\Phi|^2-1)
+
\sin\omega t\cos\omega t\, x_2\partial_1(|\Phi|^2-1)
\ccr
&+
\sin\omega t\cos\omega t\,x_1\partial_2(|\Phi|^2-1)
+
\sin^2\omega t\, x_2\partial_2(|\Phi|^2-1)\Big)d^2\vec{x}.
\cr}
\eqno(A.3)
$$
Integrating by parts yields now, 
{\it using that $|\Phi|\to1$ at infinity}, yields the
last  relation in (3.18), with 
 the particle number $N$ being defined by
Eq. (2.10). It is worth mentionning that
these commutation
 relations are in fact the remnants of those of the
centrally extended Galilei group.

\goodbreak
\vskip5mm\goodbreak
\centerline{\bf References}
\vskip-1mm
\reference
N. Manton, 
Ann. Phys. (N. Y.). {\bf 256}, 114 (1997).

\reference 
M. Le Bellac and J.-M. L\'evy-Leblond,
{\sl Il Nuovo Cimento} {\bf 14B}, 217 (1973).

\reference
R.~Jackiw and S-Y.~Pi, 
{\sl Phys. Rev. Lett}. {\bf 64}, 2969 (1990);
{\sl Phys. Rev}. {\bf D42}, 3500 (1990).
For reviews, see
R. Jackiw and S-Y. Pi,
{\sl Prog. Theor. Phys. Suppl}. {\bf 107}, 1 (1992),
or
G. Dunne, {\sl Self-Dual Chern-Simons Theories}.
Springer Lecture Notes in Physics. New Series: Monograph 36. (1995).

\reference 
A. A. Abrikosov, {\sl Sov. Phys. JETP} {\bf 5}, 1174 (1957).
The Abrikosov vortices have been imbedded into the
relativistic model of H. B. Nielsen and P. Olesen, 
{\sl Nucl. Phys}. {\bf B61}, 45 (1973).

\reference
E. B. Bogomolny, 
{\sl Sov. J. Nucl. Phys}. {\bf 24}, 449 (1976);
H. J. De Vega and F. A. Schaposnik, 
{\sl Phys. Rev}. {\bf D14}, 1100 (1976).

\reference 
E. Weinberg, {\sl Phys. Rev}. {\bf D19}, 3008 (1979);
C. H. Taubes,
{\sl Commun. Math. Phys}. {\bf 72}, 277 (1980).

\reference
I. Barashenkov and A. Harin 
{\sl Phys. Rev. Lett}. {\bf 72}, 1575 (1994);
{\sl Phys. Rev}. {\bf D52}, 2471 (1995).

\reference 
P. Forg\'acs and N. Manton, 
{\sl Commun. Math. Phys.} {\bf 72}, 15 (1980).
A similar definition has appeared in 
R.~Jackiw, {\sl Phys. Rev. Lett}. {\bf 41}, 1635 (1979).
The associated conservation laws are studied in 
R. Jackiw and N. Manton, 
{\sl Ann. Phys}. (N. Y.) {\bf 127}, 257 (1980).
See also R. Jackiw, 
{\sl Acta Physica Austr}. (Suppl.) {\bf 22}, 383 (1980).

\reference 
Z. F. Ezawa, M. Hotta and A. Iwazaki,
{\sl Phys. Rev. Lett}. {\bf 67}, 411 (1991),
{\sl Phys. Rev}. {\bf D44}, 452 (1991).
See also R.~Jackiw and S-Y.~Pi,
{\sl Phys. Rev. Lett}. {\bf 67}, 415 (1991);
{\sl Phys. Rev}. {\bf D44}, 2524 (1991);
M. Hotta {\sl Prog. Theor. Phys}. {\bf 86}, 1289 (1991).

\reference 
C. Duval, P. A. Horv\'athy and L. Palla,
{\sl Phys. Rev}. {\bf D50}, 6658 (1995).

\reference 
U. Niederer, {\sl Helvetica Physica Acta} {\bf 46}, 191 (1973).

\reference  
C. Duval, P. A. Horv\'athy and L. Palla,
{\sl Phys. Rev}. {\bf D52}, 4700 (1995);
Ann. Phys. (N. Y.) {\bf 249}, 265 (1996).

\reference
C. Lee, K. Lee, H. Min, 
{\sl Phys. Lett}. {\bf B252}, 79 (1990).

\reference
K. Lee,  
{\sl Phys. Rev}. {\bf D49}, 4265 (1994);
K. Lee and P. Yi, 
{\sl Phys. Rev}. {\bf D52}, 2412 (1995).

\reference
 P. Donatis and R. Iengo, 
 {\sl Phys. Lett}. {\bf B320}, 64 (1994).

\reference
S. K. Paul and A. Khare, 
{\sl Phys. Lett}. {\bf B174}, 420 (1986).

\reference
B. Julia, A. Zee, 
{\sl Phys. Rev. Lett}. {\bf 11}, 2227 (1975).

\reference
G. Dunne and C. Trugenberger, 
{\sl Phys. Rev}. {\bf D43}, 1323 (1991).

\reference %
L. Martina, O. K. Pashaev, G. Soliani,
{\sl Phys. Rev}. {\bf B48}, 15 787 (1993),
Z. N\'emeth, Budapest preprint (1997).

\reference  
C. Duval, P. A. Horv\'athy and L. Palla,
{\sl Phys. Lett}. {\bf B325}, 39 (1994).

\reference 
S. M. Girvin in {\sl The Quantum Hall Effect},
edited by R. E. Prange and S. M. Girvin (Springer Verlag,
N. Y. 1986), Chapt. 10.
See also S. M. Girvin and A.-H. Mac Donald, 
{\sl Phys. Rev. Lett}. {\bf 58}, 303 (1987).
\end